\documentstyle[12pt,epsfig]{article}
 \newcommand{\be}{\begin{eqnarray}}
 \newcommand{\ee}{\end{eqnarray}}
 \newcommand{\beq}{\begin{equation}}
 \newcommand{\eeq}{\end{equation}}
 \newcommand{\ba}{\begin{array}{1}}
 \newcommand{\ea}{\end{array}}
 \newcommand{\bb}{\begin{thebibliography}}
 \newcommand{\eb}{\end{thebibliography}}

 \newcommand{\abstitle}[1]{{\small {\bf #1}}}
 \newcommand{\absauthor}[1]{{\small {\bf #1}}}
 \newcommand{\address}[1]{{\it #1}}
 
\begin{document}
 \begin{center}
   \abstitle{Proton distributions in $dp$ dielectron production within Regge 
     theory}
 \vspace{0.3cm}
 \absauthor{
%B.Friman$^{a}$, 
A.P.Jerusalimov, G.I.Lykasov }\\ [2.0mm]
 \address{
%$^{a}$GSI, Darmstadt, Germany, \\
JINR, Dubna, Moscow region, 141980, Russia }
\end{center}
% \vspace{0.1cm}
%Version of May 30, 2008\\
 \vspace{0.2cm} 
%{\bf Abstract}
\vskip 5mm
\begin{abstract}
The processes of dielectron production in deuteron-proton
collisions at intermediate incident deuteron beam energies are analyzed in 
the spectator model within the one-pion exchange reggeized approach. 
We focus mainly on the momentum and angle distributions of the proton-spectator 
and the proton emitted in quasi-free $NN$ processes at small 
angles in the laboratory frame. It is shown that the inclusion 
of many channels in quasi-free $NN$ interaction allows us to describe the 
HADES data quite satisfactorily at incident deuteron kinetic energies of
about 2.5 GeV. 

\end{abstract}

%\vspace{0.1cm}
 
\vspace{1cm}\vspace{1cm}
\section{Introduction}
%%%%%%%%%%%%%%%%%%%%%%%% HS'15 %%%%%%%%%%%%%%%%%%%%%%%%%%%%%%%%%

As it is well known, the Regge theory \cite{Collins} is well applied in the analysis of 
inclusive hadron production in nucleon-nucleon and pion-nucleon collisions at high energies 
and not large transfers. For analysis of the exclusive production of one or two pions 
in $NN$ or $\pi N$ interactions the one-pion exchange reggeized
(OPER) model was suggested   
\cite{Ponomarev}. 
%This approach was succesfully applied to describe rather satisfactorily the 
%momentum spectra of produced pions or nucleons   in $NN$ or $\pi N$ collisons at intial energies 
%starting from a few GeV. 
For example, recently the OPER was successfully applied \cite{pipiN,HADES:2015_np} 
to describe the experimental data on two-pion production in $np$ collisions at incident 
kinetic energies 
$E_{kin}$ of about 1-5 GeV \cite{pipiN,OPER,OPER2,JERUS:2015} and the HADES data at $E_{kin}=$ 1.25 GeV 
\cite{HADES:2015_np}.   
One of the most interesting goals in the HADES program is the study 
of $e^+ e^-$-pair production in $np$ interactions. But there are no pure narrow 
neutron beams in the world. Thus, this process can only be studied 
using the $dp$ or $pd$-interactions.                           

The HADES~\cite{HADES} ({\bf H}igh {\bf A}acceptance {\bf D}i-{\bf E}electron 
  {\bf S}pectrometer) is designed at GSI (Darmstead, Germany) and operated at the   
  SIS synchrotron at nuclear ($A$) beam energies of about 1-2 A GeV. It is a magnetic 
  spectrometer capable of registering both hadrons (p, K- and $\pi$-mesons) and 
$e^+$/$e^-$ pairs within the range of polar angles from $17^{\circ}$ up to 
$85^{\circ}$ and exhibits a nearly full azimuth coverage. The main goal of the HADES 
experiments is the study of hadron properties inside the hot and dense nuclear 
medium via their di-electron decays. One specific aspect of heavy-ion reactions 
in the 1-2 AGeV range is the important role of baryonic resonances produced, 
that propagate due to the long lifetime of the dense hadronic 
matter phase. A detailed description of resonance excitation and its coupling to 
the pseudo-scalar and vector mesons is significant for interpretation of the 
di-electron spectra measured by HADES. 

The broad experimental program includes the study of $e^+ e^-$-pair production 
in nucleus-nucleus collisions, elementary reactions ($pp$, $np$, $\pi p$) as well 
as $pA$, $\pi A$ collisions with the emphasis on properties of vector mesons at 
finite baryonic densities. 
\section{Calculation procedure}
\subsection{Spectator mechanism for dielectron production in the $d p$ reaction}

\begin{figure}[h]
\hspace{4.0cm}
\centerline{\includegraphics[width=12.0cm]{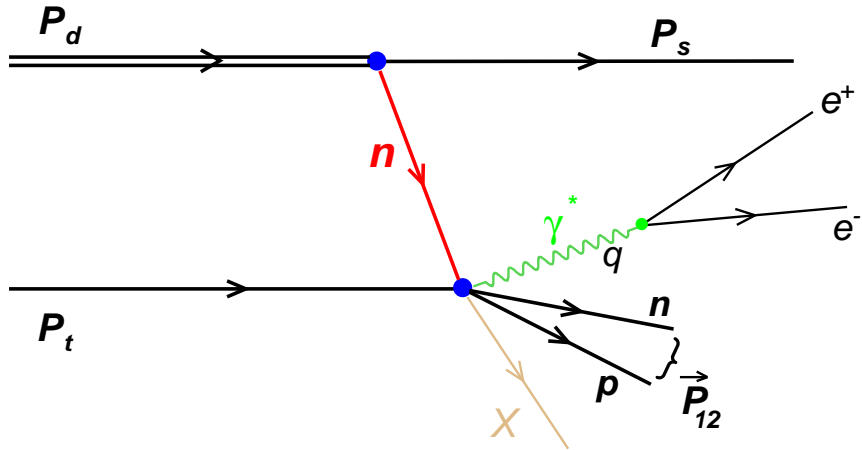}}
%diag4.eps}}
\caption{The spectator diagram for the process $dp\rightarrow p_s e+ e- X$. 
  Here  $P_d, P_t, P_S$ and $\vec P_{12}$ stand for the momenta of 
  the deuteron, proton-target, proton-spectator 
  and of the three-momentum of the final proton ($p$) and neutron ($n$), 
  respectively; 
$X$ represents one or two pions.   
} 
\label{dp_spmod}
\end{figure}

If the proton is emitted in the forward direction, i.e., at small angles relative to the 
beam in the 
reaction $d p\rightarrow p e^+ e^- X$, then one can use the so-called spectator diagram 
to analyze spectra of the hadrons and dielectrons 
produced. Here the system $X$ can contain two nucleons and a few pions. 
This diagram is presented in 
Fig.~\ref{dp_spmod}, where $p_s$ is the spectator proton. 
In this case the dielectron pair is produced in the quasi-free 
$n p\rightarrow e^+ e^- X$ reaction. 
In principle, the neutron can also be treated as a spectator, ($n_s$), then
dielectrons are produced in the quasi-free 
$p p\rightarrow e^+ e^- X$ process.   
To resolve the spectator neutron problem the HADES set-up \cite{HADES} 
was upgraded with the {\bf F}orward scintillator hodoscope {\bf W}all 
({\bf FW}) ~\cite{FW_AngMee,HADES:2014} to select quasi-free $n-p$ reactions by detecting fast spectator  
protons from the deuteron break-up. The {\bf FW} was located 7 m downstream from  
the proton target covering polar angles between $0.3^{\circ}$ and $7.0^{\circ}$ and provided 
the time-of-flight information that permitted to reconstruct the proton-spectator momentum.

The {\bf FW} can register not only the proton-spectator, but also other charged  
particles (mainly secondary protons). 
%As for a spectator the neutron can be also treated, which is not observed,  and 
%%The another source of the background are 
%the charged secondary particles from the quasi-free $pp$-interaction 
%are produced. 

To eliminate dilepton pairs produced by $\gamma$ conversion
in the detector material the $e^+ e^-$-pairs were selected at angles
$\Theta^{LAB}_{ee}>9^{\circ}$.

The spectra of nucleons and of dielectron pairs produced in the $dp$ reaction  
were calculated by MC simulation inputting the nucleon momentum  
distribution inside the deuteron and the differential cross section of the processes 
$NN\rightarrow NN X e^+ e^-$. As for the deuteron wave function (DWF), the CD-Bonn DWF 
\cite{CD_Bonn} was used. The differential cross section of the $NN\rightarrow e^+ e^- X$ reaction 
was calculated within the one-pion exchange reggeized (OPER) model \cite{Ponomarev}.  
%----------------- 
\subsection{OPER}

\begin{figure}[h]
\hspace{4.0cm}
\centerline{\includegraphics[width=7.0cm]{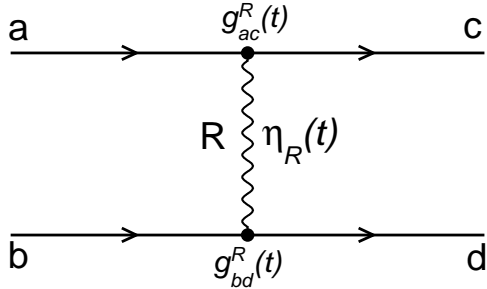}}
\caption{Diagram of the process $a+b \rightarrow c+d$}
\label{Regge}
\end{figure}

The OPER model is based on the method of  complex momenta \cite{Collins}. 
%and consider  an exchange in t-channel by a virtual
%state R that has quantum numbers of particle (resonances) with variable spin and
%is on some trajectory $\alpha_R(t)$ calle as the Regge trajectory. 
In Fig.~\ref{Regge} the one-Reggeon $R$ exchange graph in the $t$ channel of
the binary process 
$a+b\rightarrow c+d$ is presented.
The virtual state R has quantum numbers of a particle or a 
resonance with a certain spin and orbital momentum in the 
complex space, which corresponds to a Regge trajectory $\alpha_R(t)$ 
that is a function of the transfer $t$. The 
amplitude of  binary processes $a+b \rightarrow c+d$ presented in Fig.~\ref{Regge}
is given by the following \cite{Collins}:
\begin{equation}
\label{ReggeEx}
  T_R(s,t)=i8 \pi s_0 \; g_{ac}^R(t) \; \eta_R(t) \left( \frac{s s_0}
  {m_c^2 m_d^2} \right)^{\alpha_R(t)} \!\! g_{bd}^R(t)~.
\label{def:TR}
\end{equation}
Here $s$ is the initial energy squared, $s_0$ is the energy scale factor, which is usually chosen 
to be about 1 GeV$^2$, $m_c,m_d$ 
are the masses of the produced hadrons $c$ and $d$, respectively; 
%\hspace*{1.7cm} 
$g_{ac}^R(t)$ ( $g_{bd}^R(t)$) is the vertex function or the so-called Regge form 
factor depending on the transfer $t$,  
which is usually parametrized in the simple exponential form 
$g_{ac}^R(t)=g_{ac}^R(t=0)exp(-R^2\mid t\mid)$, where the parameter $R$ is called the Regge 
radius, 
%\hspace*{1.7cm} 
$\eta_R(t)$ is the signature factor, determined as follows: 
\begin{equation}
 \eta_R(t) = - \frac{\sigma + \exp(-i \pi \alpha_R(t))}{\sin{[\pi \alpha_R(t)]}}~, 
\label{def:sigfact}
\end{equation}
where $\sigma=\pm$1 is the signature value. If $\sigma=+$1, then   
$\eta_R(t)=i-ctg(\pi \alpha_R(t)/2)$ and $\eta_R(t)=i-tg(\pi \alpha_R(t)/2)$  at $\sigma=-$1.
 
The pion Regge trajectory, which is a linear function of $t$, can be expanded
in the following approximate form, see for example
\cite{Kamalov:1976ui,Beketov:1979rm}:
\begin{equation}
\alpha_R(t)~=~\alpha_R(t=0)~+~\alpha^\prime_R(t=0)\cdot t, 
\label{def:rtraj}
\end{equation}
where $\alpha_R(t=0)$  is the intercept of the Regge trajectory, $\alpha^\prime_R(t=0)$ is its
derivative at $t=0$.
\begin{figure}[h]
\hspace{4.0cm}
\vspace*{8pt}
\centerline{\includegraphics[width=14.0cm]{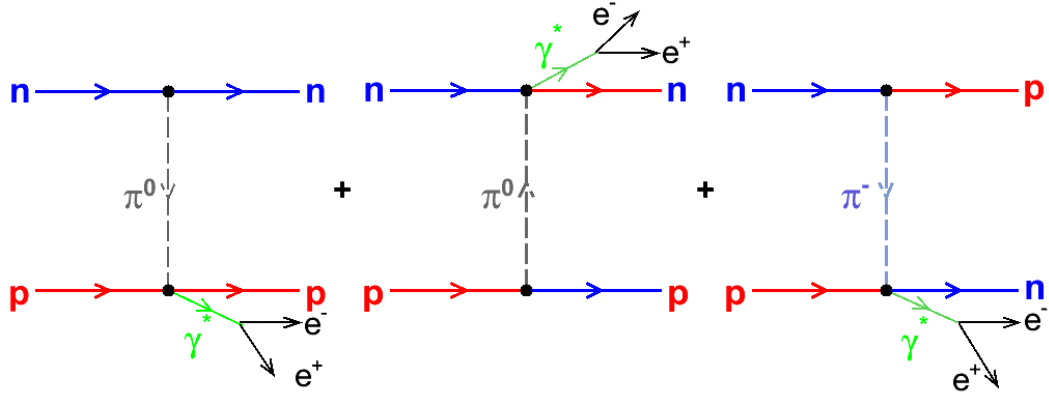}}
\caption{The one-pion exchange reggeized for the reaction 
$n p\rightarrow\gamma* n p\rightarrow e^+e^- n p$}.
\label{np_diel}
\end{figure}
The diagrams of the one-pion exchange reggeized for the reaction 
$n p\rightarrow\gamma* n p\rightarrow e^+e^- n p$ 
are presented in Fig.~\ref{np_diel}. The amplitude of this process within the OPER can  be
presented in the following form: 
\begin{equation}
T_{NN \rightarrow NN \gamma^* \rightarrow NN e^+e^-}=
  G_{\pi N} \bar{\nu(p_N')} \Gamma_{\pi N} u(p_N) F_{\pi N}
  f_{\pi N \rightarrow \gamma^* N \rightarrow e^+e^- N}
\label{def:npdielR}
\end{equation}
where 
%\hspace*{1.1cm}where 
\begin{equation}
F_{\pi N}=
 exp[(R^2_1+\alpha_{\pi}'ln(s(\mu^2+k^2_{\gamma^* \bot})/(s_0s_1)))(t-\mu^2)]
\label{def:FFR}
\end{equation}
Here, the following notation is introduced:  
 $\Gamma_{\pi N}=\gamma_5 q_m \gamma^m$ or $\Gamma_{\pi N}=\gamma_5$; 
 $s=(p_n+p_p)^2$; $s_1=(p_{\pi}+p_p)^2$; $s_2=(p_n+k_{\gamma^*})^2$; 
 $t=(p_N'-p_p)^2$; $p_n,p_p,p_\pi$ are the four-momenta of the initial neutron, proton and  
exchange pion, respectively; $p_N,p^\prime_N,k_{\gamma*}$ are the four-momenta of
the final nucleon and virtual time-like photon decaying into
a $e^+e^-$ pair.
  
The slope of the pion trajectory $\alpha_{\pi}'=$0.7 is well-known, see, for 
example, \cite{Ponomarev}. The parameter value $R^2_1=$3.3 GeV$^{-2}$ is chosen from the best 
description of many observables of
dielectron production in the quasi-free $np$ interaction presented below. 
In the general case we calculated the amplitude of the process
$NN\rightarrow\gamma* n p\rightarrow e^+e^- X$, where $X$ can stand for 
two nucleons and one or two pions. In the next subsection we present the different channels 
of dielectron production in the $NN$ interaction.

\subsection{Different channels of  dielectron production in quasi-free $NN$
                  interactions}
\label{res}

  The following reactions were taken into account to describe the production of 
  $e^+ e^-$-pairs ($p_s$ denotes the spectator proton and $n_s$ is the
  spectator neutron:\\
\hspace{1.0cm} $dp \rightarrow p_s+(np \rightarrow e^+ e^- +X)$:\\
\hspace*{3.1cm} $np \rightarrow np e^+ e^-$,\\
\hspace*{3.1cm} $np \rightarrow np \pi^0$, 
   \hspace{5mm} $\pi^0 \rightarrow e^+ e^- \gamma$\\
\hspace*{3.1cm} $np \rightarrow np \pi^0 \pi^0$, 
   \hspace{1mm} $\pi^0 \rightarrow e^+ e^- \gamma$\\
\hspace*{3.1cm} $np \rightarrow \Delta^0 p$,
   \hspace{6mm} $\Delta^0 \rightarrow n e^+ e^-$\\
\hspace*{3.1cm} $np \rightarrow \Delta^+ n$,
 \hspace{4.2mm} $\Delta^+ \rightarrow p e^+ e^-$\\
\hspace*{3.1cm} $np \rightarrow \Delta^0 N \pi$,
   \hspace{2mm} $\Delta^0 \rightarrow n e^+ e^-$, 
                $\quad N \pi = n \pi^+ \; \& \; p \pi^0$\\
\hspace*{3.1cm} $np \rightarrow \Delta^+ N \pi$,
 \hspace{0.5mm} $\Delta^+ \rightarrow p e^+ e^-$,
                $\quad N \pi = n \pi^0 \; \& \; p \pi^-$\\
\hspace*{3.1cm} $np \rightarrow np \eta^0$, 
   \hspace{6mm} $\eta^0 \rightarrow e^+ e^- \gamma$\\
\hspace*{3.1cm} $np \rightarrow np \rho^0$, 
   \hspace{6mm} $\rho^0 \rightarrow e^+ e^-$\\

\hspace{0.35cm} $dp \rightarrow n_s+(pp \rightarrow e^+ e^- +X)$:\\
\hspace*{3.1cm} $pp \rightarrow pp \pi^0$, 
   \hspace{5mm} $\pi^0 \rightarrow e^+ e^- \gamma$\\
\hspace*{3.1cm} $pp \rightarrow pp \pi^0 \pi^0$, 
   \hspace{1mm} $\pi^0 \rightarrow e^+ e^- \gamma$\\
\hspace*{3.1cm} $pp \rightarrow \Delta^+ p$,
 \hspace{3.8mm} $\Delta^+ \rightarrow p e^+ e^-$\\
\hspace*{3.1cm} $pp \rightarrow \Delta^0 p \pi^+$,
   \hspace{0mm} $\, \Delta^0 \rightarrow n e^+ e^-$,\\ 
\hspace*{3.1cm} $pp \rightarrow \Delta^+ N \pi$,
   \hspace{0mm} $\Delta^+ \rightarrow p e^+ e^-$, 
                $\quad N \pi = p \pi^0 \; \& \; n \pi^+$\\

The rates of these processes to the total yield of the $np\rightarrow npe^+e^-X$ 
at the effective dielectron mass $M_{ee} < 0.140 GeV/c^2$ are presented in Table 1.

\begin{table}[h]
{
\baselineskip 18pt
\caption{$M_{ee} < 0.140 GeV/c^2$}
\label{tb1}
\begin{center}
\baselineskip 10pt
\begin{tabular}{|l|c|c|c|}

\hline
 \bf Reaction&$0.5^{\circ} < \Theta < 2.0^{\circ}$&
              $2.0^{\circ} < \Theta < 4.0^{\circ}$&
              $4.0^{\circ} < \Theta < 6.0^{\circ}$\\
\hline
% \bf ${\color{blue}n}{\color{red}p} \rightarrow np \pi^0$         &0.840   %&0.765&0.538\\
 \bf $np \rightarrow np \pi^0$         &0.826   &0.759&0.533\\
\hline
 \bf $np \rightarrow np \pi^0 \pi^0$   &0.036   &0.064&0.063\\
\hline
 \bf $np \rightarrow np e^+ e^-$       &0.041   &0.005&0.004\\
\hline
 \bf $np \rightarrow \Delta N$         &0.001   &0.001&0.001\\
\hline
 \bf $np \rightarrow \Delta N \pi$     &0.025   &0.013&0.011\\
\hline
 \bf $np \rightarrow np \eta^0$        &0.017   &0.008&0.007\\
\hline
\hline
 \bf $pp \rightarrow pp \pi^0$         &0.045   &0.118&0.212\\
\hline
 \bf $pp \rightarrow pp \pi^0 \pi^0$   &0.002   &0.003&0.005\\
\hline
 \bf $pp \rightarrow \Delta^+ p$       &$<$0.001&0.001&0.004\\
\hline
 \bf $pp \rightarrow \Delta^+ N \pi$   &0.007   &0.028&0.160\\
\hline

\end{tabular}
\end{center}
}
\end{table}
One can see from Table 1 that the main contribution to dielectron production
in the $np$ 
interaction at $M_{ee}<$ 0.14 GeV$/$c$^2$ comes from the channel $np\rightarrow np\pi^0$ 
with the subsequent $\pi^0\rightarrow e^+e^-\gamma$ decay. The
decay amplitudes of
$\pi^0,\rho^0,\eta^0$-mesons  and the
$\Delta$-isobar were calculated using the PLUTO model \cite{dGdMee,Pluto,Pluto_MC}.

\begin{figure}[h!]
\centerline{\includegraphics[width=14.0cm]{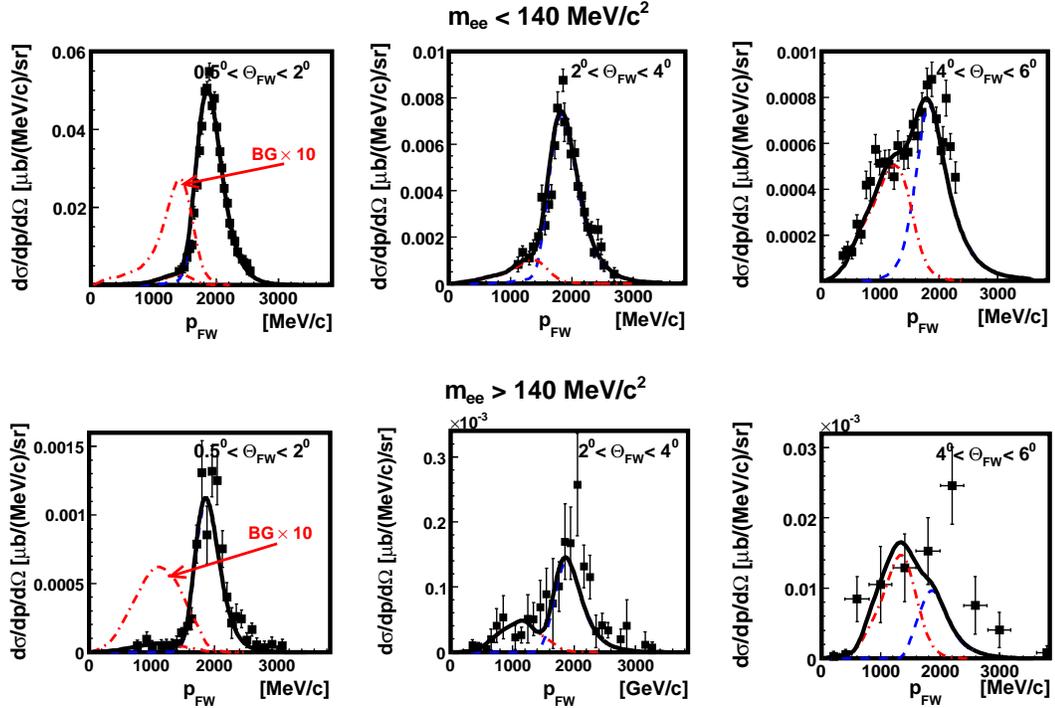}}
%{PspDistp1.eps}}
\vspace*{8pt}
\caption{The momentum spectra of protons registered by the {\bf FW} 
within two ranges of di-electron masses and three ranges of $\Theta_{FW}$.  
Black squares are experimental data \cite{Pspect}, the solid line 
represents the total description, the    
dashed curve is the contribution ofm a true spectator proton, 
the dash-dotted line corresponds to the background, when the proton produced in 
reaction $Np\rightarrow p X$ is registered by the {\bf FW}.} 
\label{Psp}
\end{figure}

The rates of these processes to the total yield of the $np\rightarrow npe^+e^-X$
at the effective dielectron mass $M_{ee} > 0.140 GeV/c^2$ are presented in Table 2.
\begin{table}[h]
{
\baselineskip 18pt
\caption{$M_{ee} > 0.140 GeV/c^2$}
%\hspace{1.0cm}
\label{tb1}
\begin{center}
\baselineskip 10pt
\begin{tabular}{|l|c|c|c|}

\hline
 \bf Reaction&$0.5^{\circ} < \Theta < 2.0^{\circ}$&
              $2.0^{\circ} < \Theta < 4.0^{\circ}$&
              $4.0^{\circ} < \Theta < 6.0^{\circ}$\\
\hline
 \bf $np \rightarrow np e^+ e^-$       &0.463   &0.292&0.259\\
\hline
 \bf $np \rightarrow \Delta N$         &0.146   &0.127&0.138\\
\hline
 \bf $np \rightarrow \Delta N \pi$     &0.104   &0.110&0.137\\
\hline
 \bf $np \rightarrow np \eta^0$        &0.139   &0.195&0.211\\
\hline
 \bf $np \rightarrow np \rho^0$        &0.032   &0.032&0.037\\
\hline
\hline
 \bf $pp \rightarrow \Delta^+ p$       &0.054   &0.066&0.121\\
\hline
 \bf $pp \rightarrow \Delta^+ N \pi$   &0.062   &0.178&0.097\\

\hline

\end{tabular}
\end{center}
}
\end{table}
%-------------
One can see from Table 2 that  at $M_{ee}>$ 0.14 GeV$/$c$^2$
the main contribution to the dielectron production in the $np$
interaction comes from the channel $np \rightarrow np e^+ e^-$,
which was calculated in \cite{JL:2015}.
%\vspace*{8pt}
Fig.~\ref{Psp} presents the momentum distributions of charged particles (protons) 
calculated within the OPER including the possible channels (see table 1 and Table 2)
and their comparison to the HADES data \cite{Pspect}.
At the top of Fig.~\ref{Psp}, the distributions 
are presented as functions of the proton momentum 
$p_{FW}$ registered by the {\bf FW} at $M_{ee}<$ 140 MeV$/$c
within three intervals of the angle $\Theta_{FW}$ between the proton and the 
incident deuteron beam. 
At the bottom of Fig.~\ref{Psp}, the same 
distributions are presented for $M_{ee}>$ 140 MeV$/$c. The solid lines
in Fig.~\ref{Psp} correspond to our total calculation including the
contributions of channels
presented in Table 1 and Table 2; the dashed lines are the contributions
due to the spectator proton, the dash-dotted curves correspond
to the background, which is the contribution of protons produced in the reaction
$Np\rightarrow pX$ illustrated by the bottom vertex in Fig.~\ref{dp_spmod}. All the results
presented in Fig.~\ref{Psp} were obtained including the possible $np$ and $pp$
channels listed in Tables (1,2). 
Note that the channels
$pp \rightarrow pp \pi^0,pp \rightarrow pp \pi^0 \pi^0,pp \rightarrow \Delta^+ p, 
pp \rightarrow \Delta^+ N \pi$
correspond  to the contribution of the spectator 
neutron. One can see from Fig.~\ref{Psp} that the background contribution is visible in the 
$p_{FW}$-spectrum only at large angles $2.0^{\circ} < \Theta_{FW} < 4.0^{\circ}$
and it is sizable at $4.0^{\circ} < \Theta_{FW} < 6.0^{\circ}$.  
The description of the HADES data within the OPER and one-baryon exchange (OBE)
\cite{OBE} model is satisfactory except in the region of 
$4.0^{\circ} < \Theta_{FW} < 6.0^{\circ}$ at $M_{ee}>$140 MeV and at 
$p_{FW}>2000$ MeV$/$c, where the error bars are large. 

\begin{figure}[h!]
\centerline{\includegraphics[width=14.0cm, height=6.0cm]{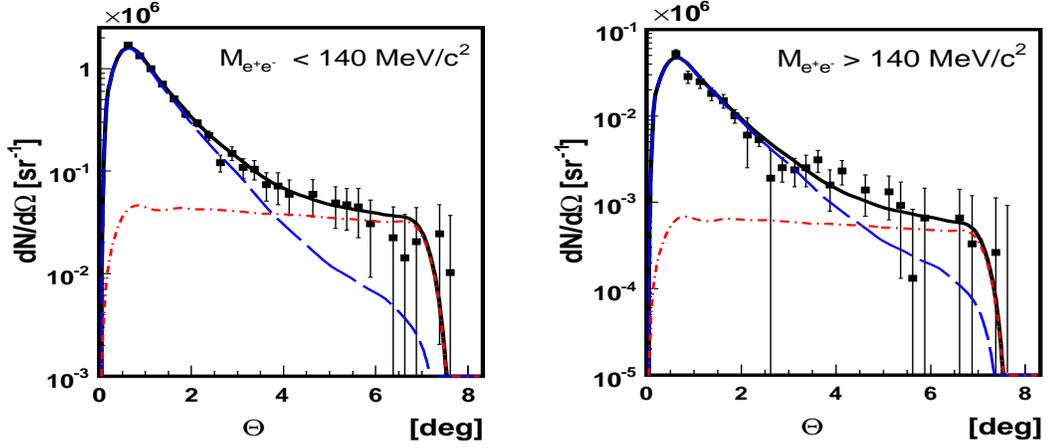}}
%{AngPFWlc.eps}}
%{AngPFWlp1.eps}}
\vspace*{8pt}
\caption{The angular spectra of the charged particles registered at the {\bf FW}. 
  Black squares are the experimental data \cite{Pspect}, the solid line is the 
  total description, the dashed line is the contribution of a true 
spectator proton, the dash-dotted 
line corresponds to the background, when the proton produced in 
the reaction $Np\rightarrow p X$ is registered by the {\bf FW}.} 
\label{AngPFW}
\end{figure}

%-----------------
Fig.~\ref{AngPFW} presents the angular distributions of protons
registered by the {\bf FW}. The experimental data were taken from~\cite{Pspect}.
The dashed line is the contribution of the spectator proton registered by the
{\bf FW}, the dash-dotted curve corresponds to the background, which is due
to the non spectator protons registered by the {\bf FW}.  
One can see the description of the HADES data to be satisfactory within the 
OPER+one-boson exchange (OBE) model. Unfortunately, at $M_{ee}>$140 MeV$/$c
and $4.0^{\circ} < \Theta_{FW} < 6.0^{\circ}$ the statistics is very poor.

\begin{figure}[h!]
\centerline{\includegraphics[width=14.0cm]{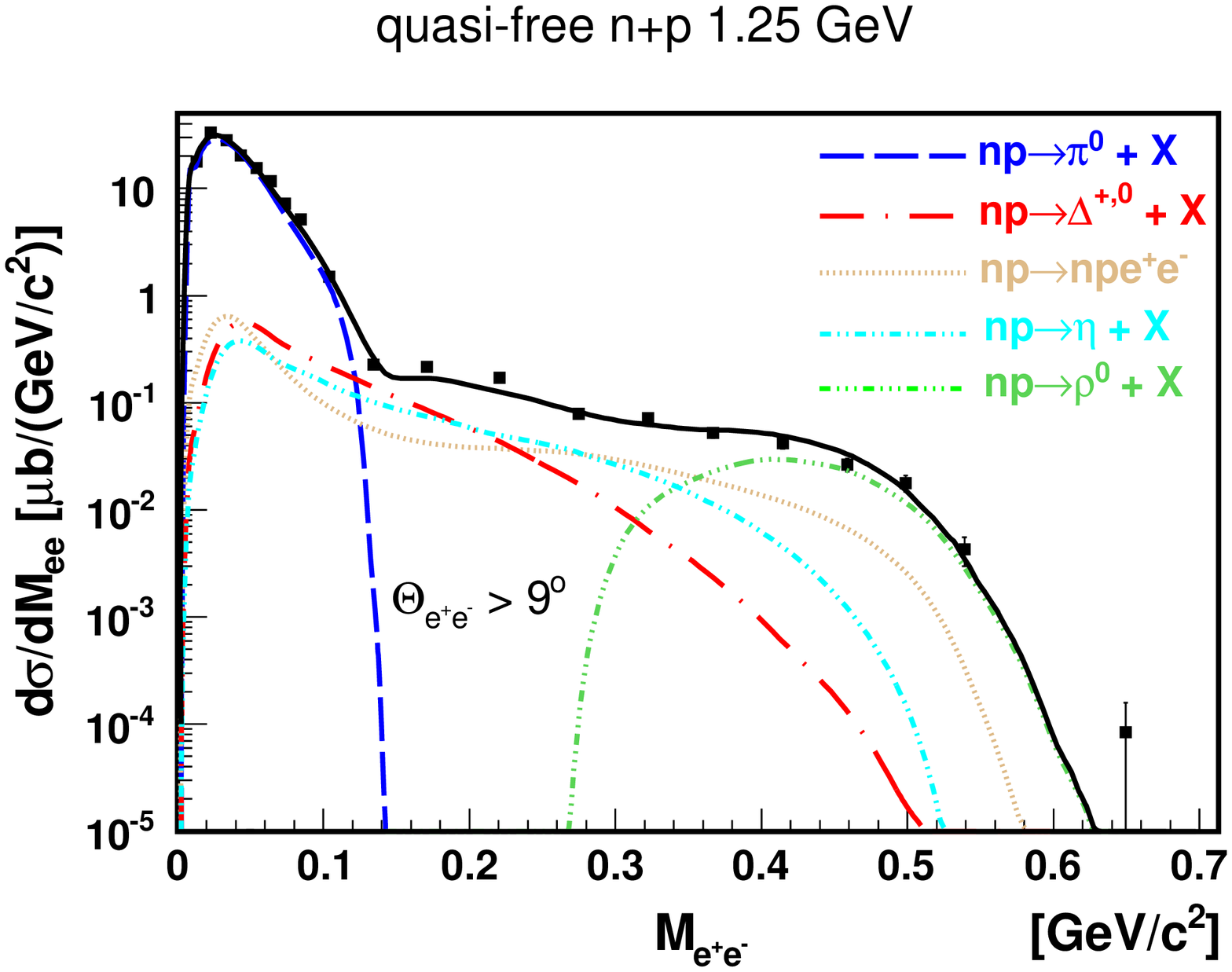}}
%{EfMeeb}}
\vspace*{8pt}
\caption{The effective mass spectrum of $e^+ e^-$-pairs. 
 The black squares represent experimental the smooth curves result from calculations 
  using the (OPER+OBE)-model involving the contributions of
  different $np$ channels. }
\label{EfMee}
\end{figure}

%-----------------
\vspace*{8pt}
 Fig.~\ref{EfMee} presents the effective mass
distribution of $e^+ e^-$-pairs produced in the quasi-free $np$ reaction. 
To achieve a satisfactory description of the HADES data
the following reactions were taken into account
~\cite{FW_AngMee,Galatyuk}:\\ 
\hspace*{0.5 cm} $d p \rightarrow p_s + (n p \rightarrow e^+ e^- + X)$\\
\hspace*{2.35 cm}$n p \rightarrow n p  \rho^0$, ($\rho^0 \rightarrow e^+ e^-$)\\
\hspace*{2.35 cm}$n p \rightarrow n p  \eta^0$, ($\eta^0 \rightarrow e^+ e^- \gamma$)

The matrix elements squared of the $\rho^0$ and $\eta^0$ Dalitz decays were calculated
according go the PLUTO model, see \cite{Pluto,Pluto_MC}.
%-----------------
One can see from Fig.~\ref{EfMee} that  the main contribution to the $M_{ee}$ spectrum 
comes from the reaction  $np \rightarrow\pi^0 X$ with the subsequent    
$\pi^0\rightarrow e^+e^-\gamma $ decay, whereas at large $M_{ee}>$ 0.3 GeV$/$c$^2$ 
the $\rho^0$ and $\eta$ meson production in the $np$ interaction contributes 
significantly and its inclusion 
allows us to describe the high effective mass behavior of the spectrum. 
Inclusion of the channels listed in Table 1 and Table 2 results in 
a satisfactory description of the $M_{ee}$ spectrum in the 
whole kinematical interval,  except the very high region of
$M_{ee}>$ 0.6 GeV$/$c$^2$.
 \begin{figure}[h!]
  \centerline{\includegraphics[width=14.0cm]{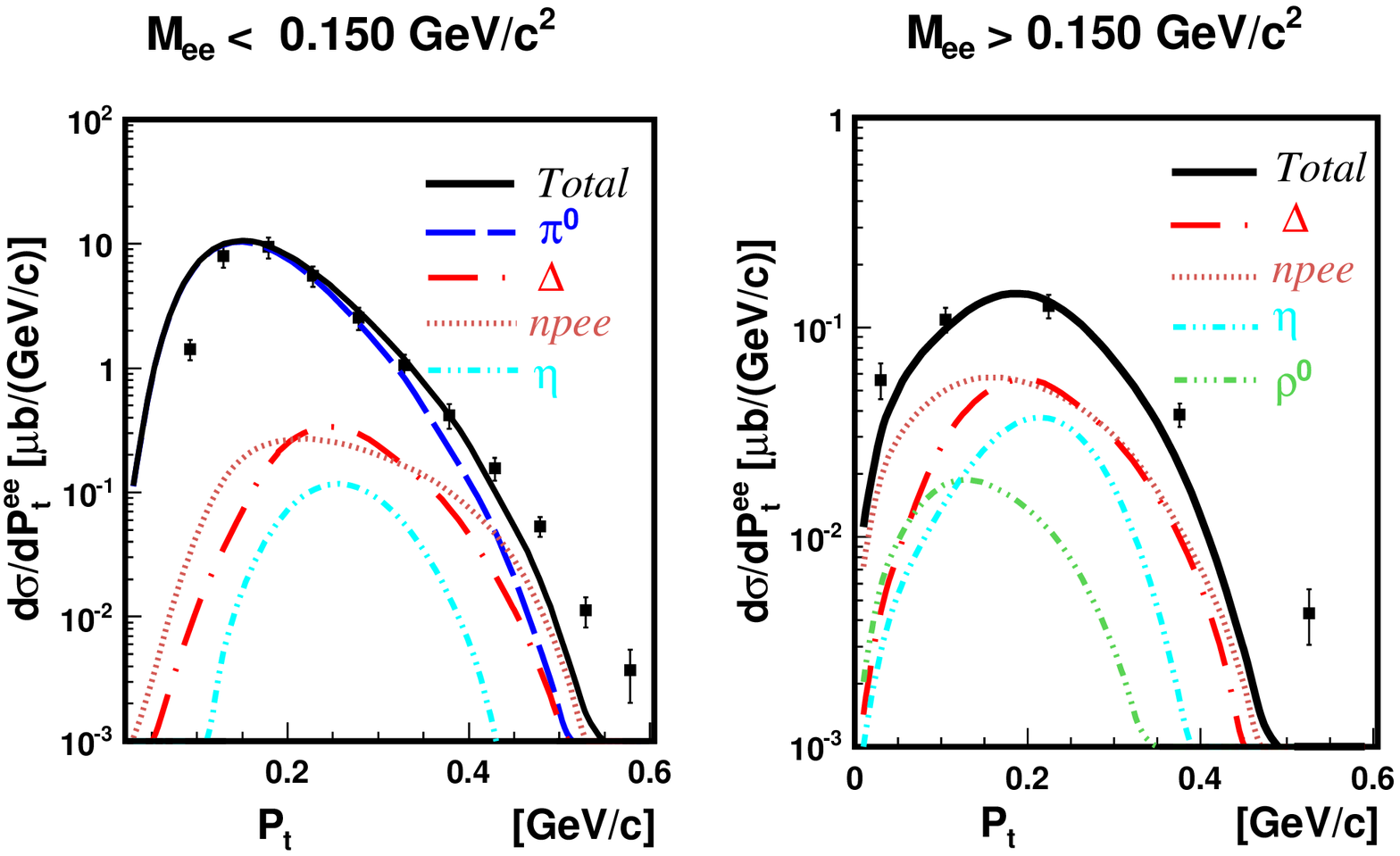}}
%{PtMee_24.02.16.eps}}
\caption{Left: the $p_T$ distribution of dielectrons at $M_{ee}\leq 0.15$ GeV$/$c$^2$.
Right: the same distribution at $M_{ee}>0.15$ GeV$/$c$^2$. 
The black squares are experimental data \cite{Galatyuk}. The solid line is our 
total calculation involving all the $np$ channels mentioned in Table 1 and
Table 2.}
\label{pTee}
\end{figure}
In Fig.~\ref{pTee} the $p_T$-spectrum of dielectrons is presented for 
 $M_{ee}\leq 0.15$ GeV$/$c$^2$ (left) and $M_{ee}>0.15$ GeV$/$c$^2$ (right). The notation of lines
 in Fig.~\ref{pTee} is the same as in Fig.~\ref{EfMee}. One can see from Fig.~\ref{pTee} that 
 inclusion of the contributions of the $np$ channels mentioned above allows us
 to describe the transverse momentum spectrum of dielectrons more or  
less satisfactorily at $p_T\leq$ 0.4 GeV$/$c. 

\section{Conclusion}
\label{con}
 
  The successful description of the proton spectra in FW (PFW) 
of HADES  was obtained by taking into account the various $np$ reactions 
and $pp$ processes, as a background. 

The reaction $np \rightarrow np\pi^0 \rightarrow npe^+e^-\gamma$  provides
the main contribution to the spectator momentum spectra at small dielectron
effective masses $M_{ee}<$ 140 MeV$/$c. 
The reaction $n p \rightarrow n p e^+e^-$  results in a significant 
contribution at M$_{ee}>$140~MeV$/$c$^2$ and     
$0.5^{\circ} < \Theta_{p,FW} < 6^{\circ}$ as compared to other channels.

At larger angles ($2.0^{\circ} < \Theta_{p,FW} < 4^{\circ}$) and 
PFW $<$ 2000 MeV/c, one should also take into account the spectator neutron  
case as well as $pp \rightarrow e^+e^- +X$  reactions. 

For a satisfactory description of the $M_{ee}$ spectrum it is 
also necessary to take into account the reactions $np \rightarrow np\eta^0$  
and $np \rightarrow np\rho^0$ with subsequent decays ($\eta^0 \rightarrow e^+e^- 
\gamma$ and $\rho^0 \rightarrow e^+e^-$  respectively) in addition to the 
reaction used for the description of PFW spectra. 

%\begin{sloppypar}
{Acknowledgements}.
The authors are grateful to members of the HADES Collaboration
P.Salabura, J.Stroth, R.Holzmann, T.Galatyuk, G.Kornakov, V.P.Ladygin, 
K. Lapidus, A.I.Malakhov, A.Rustamov, V.Pechenov for
very useful discussions, which stimulated us to do this analysis.

\end{document}